\documentclass[aps,prd,nofootinbib,superscriptaddress,10pt]{revtex4-2}
\usepackage[utf8]{inputenc}
\usepackage{graphicx}
\usepackage{subcaption}
\usepackage{amssymb,amsmath}
\usepackage{xcolor}
\usepackage{float}
 \usepackage{caption}

\def\be{\begin{equation}}
\def\ee{\end{equation}}
\def\bea{\begin{eqnarray}}
\def\eea{\end{eqnarray}}

\def\s{{\rm s}}
\def\yr{{\rm yr}}
\def\cm{{\rm cm}}
\def\km{{\rm km}}
\def\pc{{\rm pc}}
\def\kpc{{\rm kpc}}
\def\mpc{{\rm Mpc}}
\def\gev{{\rm GeV}}
\def\tev{{\rm TeV}}

\begin{document}

\title{Constraining $p$-wave dark matter annihilation with gamma-ray observations of M87}
\author{Katharena Christy}
\affiliation{Department of Physics and Astronomy, University of Hawai'i, Honolulu, Hawaii 96822, USA}

\author{Jason Kumar} 
\affiliation{Department of Physics and Astronomy, University of Hawai'i, Honolulu, Hawaii 96822, USA}

\author{Pearl Sandick}
\affiliation{Department of Physics and Astronomy, University of Utah, Salt Lake City, Utah  84112, USA}

\begin{abstract}
We consider constraints on $p$-wave dark matter in a dark matter spike surrounding the 
supermassive black hole at the center of M87.  Owing to the large mass of the black hole, and resulting 
large velocity dispersion for the dark matter particles in the spike, it is possible for Fermi-LAT 
and MAGIC data to place tight constraints on $p$-wave annihilation, which would be far more stringent than 
those placed by observations of dwarf spheroidal galaxies.  Indeed, for optimistic choices of the spike 
parameters, gamma-ray data would exclude thermal $p$-wave dark matter models with a particle mass $\lesssim 10~\tev$.  
But there is significant uncertainty in the properties and parameters of the spike, and for less optimistic scenarios, 
thermal dark matter candidates would be completely unconstrained.  In addition to better understanding the spike parameters, a second key to improving constraints on 
dark matter annihilation is an accurate astrophysical background model.
\end{abstract}

\maketitle

\section{Introduction}

If the center of a galaxy hosts a supermassive black hole (SMBH), then the 
region just outside the black hole may exhibit a large density of dark matter, known as a dark matter spike~\cite{Gondolo:1999ef}.  This region would then be a promising 
target for indirect searches for dark matter annihilation.  A significant body 
of work has been developed, focusing on searches for dark matter annihilation 
near Sgr A*, the SMBH hosted by the Milky Way (see, for 
example, \cite{Shelton:2015aqa,Sandick:2016zeg,Johnson:2019hsm,Chiang:2019zjj,Alvarez:2020fyo,Balaji:2023hmy}).  In this work, we consider the 
possibility of dark matter annihilation in a dark matter spike surrounding the 
SMBH hosted by M87, particularly for the case in which dark matter annihilates from 
a $p$-wave initial state.

The black hole at the center of M87 is interesting, from the point of view of 
dark matter searches~\cite{LBS:2015}, because it is a dynamically young galaxy.  As a result, it is
believed to be more likely that any dark matter spike at the center of M87 would have 
survived the effects of galaxy dynamics (for example, the scattering of dark matter 
against stars~\cite{Vasiliev:2008uz})~\cite{LBS:2015}.
It is also interesting because, although it is much farther away from Earth ($\sim 16~\mpc$) than 
 Sgr A* ($\sim 8.5~\kpc$), it is also about 1000 times more massive than Sgr A*, and is therefore expected to be surrounded by a more dense dark matter spike.

The size of the SMBH is especially important for the case 
of $p$-wave annihilation, because the large gravitational potential arising from 
a very massive black hole leads to a much higher velocity-dispersion for dark matter 
particles in the spike, enhancing the annihilation rate~\cite{Shelton:2015aqa}.  
Although $p$-wave annihilation 
has been studied in the dark matter spike around Sgr A* (see, for example,~\cite{Shelton:2015aqa,Sandick:2016zeg,Johnson:2019hsm}), 
we will find qualitative advantages for observations 
of M87, due to the large mass of the central SMBH, and the much greater 
likelihood that the dark matter spike in M87 has not been depleted by interactions with stars.
We note that $p$-wave dark matter annihilation in the dark matter spike in Centaurus A has also been considered, 
though in a different context~\cite{Cermeno:2022rni}.

Of course, the SMBH hosted by M87 also accretes a large amount of baryonic matter, leading to 
a variety of astrophysical processes which yield gamma-ray emission.  Indeed, gamma-ray emission 
from M87 is often classified as arising from  either a ``high-emission" or ``low-emission" state, 
due to the variability of such astrophysical emission with time.  Since the astrophysical processes 
underlying gamma-ray emission from active galactic nuclei (AGN) are only partially known, we will 
adopt a conservative bound on dark matter annihilation by assuming that all observed gamma-ray 
emission from M87 during its low-emission state arises from 
dark matter annihilation.  We focus on data from the Fermi-LAT~\cite{Fermi-LAT:2009ihh} and 
from MAGIC~\cite{Magic_the_Gathering:2020}, whose angular 
resolutions are such that M87 is essentially a point source.  

We find that, assuming the dark matter spike is not significantly depleted by 
galactic dynamics, gamma-ray data can place bounds on $p$-wave dark matter annihilation which are much 
more stringent than those which can be placed by observations of dwarf spheroidal galaxies (dSphs)~\cite{Boddy:2019qak}.  
Moreover, assuming the largest and most dense dark matter spike which is allowed by 
stellar observation, these bounds would  rule out models of 
$p$-wave thermal dark matter for masses as large as $10~\tev$.  
But the large uncertainties in the size and slope of the spike can lead to 
significant weakening of these bounds; for a small enough dark matter spike, thermal $p$-wave dark matter models would be unconstrained.

The plan of this paper is as follows.  In Sec.~\ref{sec:formalism}, we review 
the general formalism of our analysis, including a discussion of the determination of the spike profile and the calculation of the flux from annihilation of $p$-wave dark matter.  In Sec.~\ref{sec:results}, we present 
our results for constraints on the annihilation cross section and examples of differential photon fluxes for various models.  We conclude in Sec.~\ref{sec:conclusion}.

\section{General Formalism}
\label{sec:formalism}

We begin with the uncontracted M87 halo profile, as this provides the starting point which determines the form of 
the dark matter density within the central spike.  We will assume that the initial form of the 
dark matter profile is Navarro-Frenk-White (NFW), with 
$\rho(r) = \rho_s (r/r_s)^{-1} [1 + (r/r_s)]^{-2}$, where $\rho_s$ and $r_s$ are the scale density and 
scale radius, respectively~\cite{Navarro:1995iw}.  Observational studies of stellar motion can generally be used to constrain 
the gravitational potential due to dark matter, but these constraints lead to large uncertainties 
for the case of M87~\cite{Murphy:2011yz}.  We adopt the parameter estimates used in Ref.~\cite{LBS:2015}, 
namely,
$r_s = 20~\kpc$ (similar to the Milky Way halo) and $\rho_s = 2.5~\gev/ \cm^3$ (roughly an order of 
magnitude larger than for the Milky Way halo).  In the case of $s$-wave annihilation, the $J$-factor is $J \propto \rho_s^2 r_s^3$, such that we expect the absolute luminosity of 
the M87 halo due to dark matter annihilation to be roughly ${\cal O}(10^2)$ larger than that of the MW halo.

For $p$-wave annihilation within this NFW halo, the total $J$-factor is given by~\cite{Boddy:2017vpe,Boddy:2019wfg,Boucher:2021mii}
\bea
J_p^{NFW} &=& \frac{4\pi \rho_s^2 r_s^3}{D^2} \left(\frac{4\pi G_N \rho_s r_s^2}{c^2} \right) 
\tilde J_2 \sim 2 \times 10^{13}~\gev^2~\cm^{-5} ,
\eea
where $\tilde J_2 \sim 0.14$~\cite{Boddy:2019wfg,Boucher:2021mii} and $D \sim 16~\mpc$ is the distance to M87~\cite{Bird:2010rd}.  Below, we see that for profiles containing a dark matter spike the total $J$-factor will increase correspondingly.

\subsection{Spike profile}
\label{sec:spikeprofile}

The SMBH at the center of M87 is estimated to be $t_{BH} \sim 10^{10}~\yr$ old and  
has a mass of $M_{BH} \sim 6.4 \times 10^9 M_\odot$, corresponding 
to a Schwarzshild radius of $r_{sch} = 6 \times 10^{-4}~\pc$ (see, for example,~\cite{Gorchtein:2010xa}).

\begin{figure}[H]
\centering
\includegraphics[width=\textwidth]{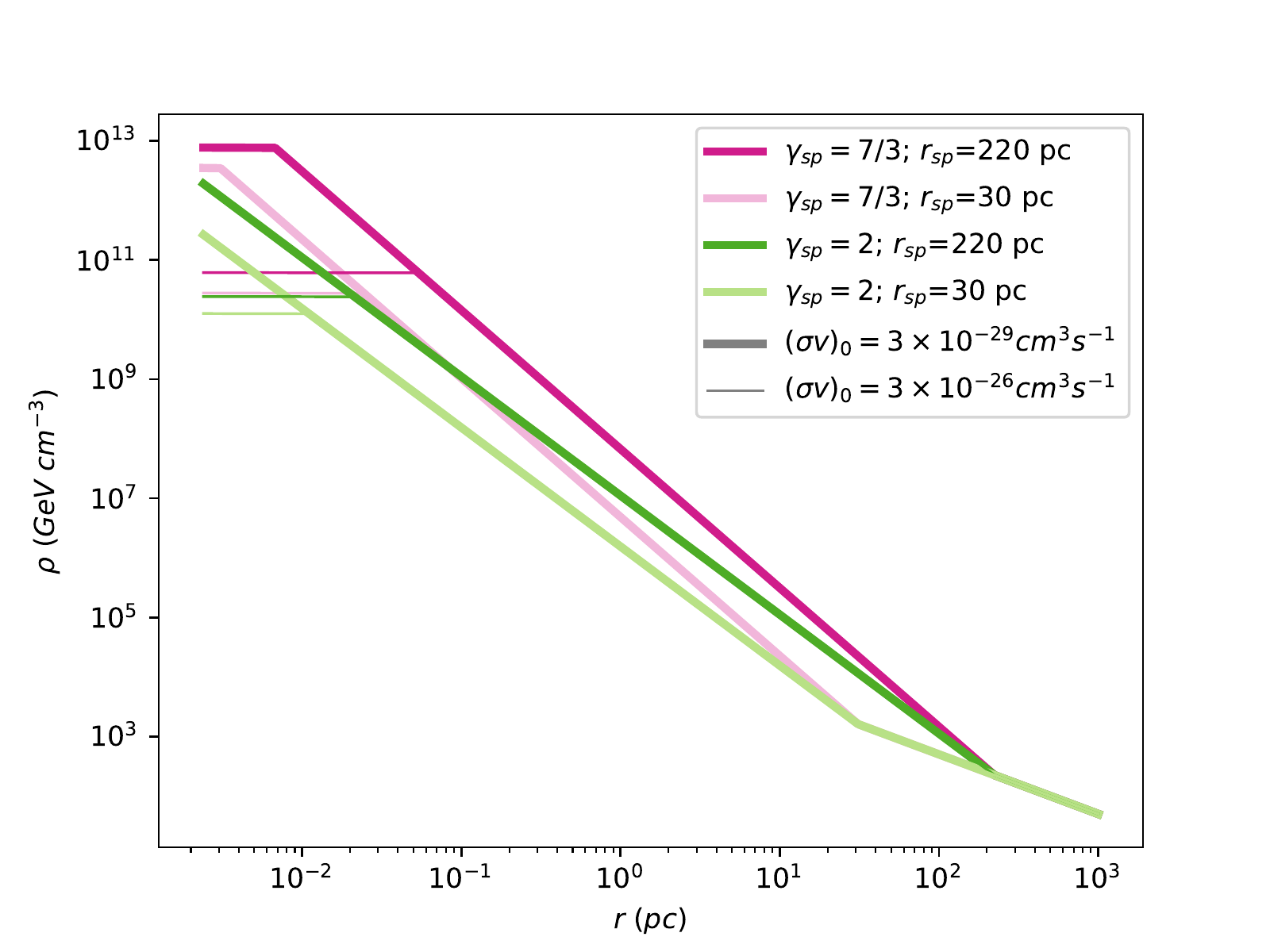}
\caption{$\rho(r)$ for $r_{sp} = 30~\pc,~ 220~\pc,$ and 
$\gamma_{sp} = 7/3$, $2$, as labeled.
Thick (thin) lines indicate $(\sigma v)_0 = 3 \times 10^{-29} \cm^3 /\s,~ (3 \times 10^{-26} \cm^3 /\s)$.
In all cases, $m=10~\gev$.}
\label{fig:profiles}
\end{figure}

Dark matter density spikes near SMBHs have been studied by many groups,
beginning with the work of Gondolo and Silk~\cite{Gondolo:1999ef}.  If the growth of the SMBH was adiabatic and dark matter particles are collisionless, one finds that the resulting dark matter density profile has four distinct regions:
\begin{itemize}
\item{$\rho(r) = 0$ for $r < r_{inner}$,}
\item{$\rho(r) = \rho_{core}$ for $r_{inner} < r < r_{core}$, }
\item{$\rho(r) = \rho_{core} \left(\frac{r}{r_{core}}\right)^{-\gamma_{sp}}$ for $r_{core} < r < r_{sp}$, }
\item{$\rho(r) = \rho_{core} \left(\frac{r_{sp}}{r_{core}}\right)^{-\gamma_{sp}} 
\left(\frac{r}{r_{sp}} \right)^{-\gamma_{c}}$ for $r_{sp} < r$. }
\end{itemize}
Here, $r_{core}$ and $r_{sp}$ are the outer radii of the core and spike regions of the dark matter profile, respectively.  
The region outside the spike corresponds to the inner slope region of a generalized NFW profile with inner-slope 
$\gamma_c$, which we take to be 1, corresponding to a standard NFW profile.  Inside the dark matter spike, which has radius $r_{sp}$, the dark matter 
density profile has a steeper slope given by the spike exponent $\gamma_{sp}$.  The dark matter density continues to grow with 
decreasing $r$ until one reaches the core radius, $r_{core}$.  Within the core, the dark matter density is so 
large that the dark matter abundance is depleted by annihilation.   We model the core as a region of 
constant density, $\rho_{core}$, such that $(\rho_{core}/m)\langle \sigma v \rangle t_{BH} =1$, where 
$m$ is the mass of the dark matter particle and $\langle \sigma v \rangle$ is the velocity-averaged 
dark matter annihilation cross section.\footnote{If $\langle \sigma v \rangle$ is small enough, this 
condition may never be satisfied.  In this case, $r_{core} = r_{inner}$, and dark matter annihilation 
never depletes the dark matter density appreciably.}  
Finally, we assume that the dark matter density is negligible 
inside an inner radius $r_{inner}$, as almost all dark matter in this region has fallen inside the black 
hole horizon.  We take $r_{inner} = 4r_{sch}$~\cite{LBS:2015,Shelton:2015aqa}, although values adopted in the literature vary from 
as low as $2r_{sch}$ (e.g.,~\cite{Alvarez:2020fyo}) to as large as $10 r_{sch}$ (e.g.,~\cite{Sandick:2016zeg}). We plot several illustrative examples of this spike profile in 
Fig.~\ref{fig:profiles}.

The dark matter density profile we have assumed is continuous outside of $r_{inner}$, but not smooth.  Other 
works in the literature have used smooth variations of this profile, but there is no known theoretically motivated 
choice for smoothly connecting the density profile in the regions described above.  As such, for simplicity, we 
will use the profile described above.  

If the spike was formed of collisionless dark matter purely through adiabatic contraction of the dark matter halo, one obtains a spike exponent~\cite{Gondolo:1999ef}
\begin{equation}
\gamma_{sp} = \frac{9-2\gamma_{c}}{4-\gamma_{c}}.
\label{eq:gamma_sp}
\end{equation}
However, the spike profile may deviate from the adiabatic expectation under different black hole growth scenarios.  Thus, 
although we fix $\gamma_c = 1$, we consider multiple choices for $\gamma_{sp}$.

Indeed, the formation and evolution of a dark matter spike depends on many factors, and it is unclear whether a spike, once formed, evolves in time.  If the spike radius does not evolve in time, we refer to the case as an ``idealized'' spike.  Gravitational effects of stars in the galactic nucleus may dampen or ``deplete'' the spike, manifesting as a reduction in the spike radius~\cite{Ullio:2001fb,Merritt:2003qk,Bertone:2005hw,Gnedin:2003rj,Ahn:2007ty}. 
For this analysis, we assume that $r_{inner}$, $\gamma_{sp}$ and the dark matter mass and annihilation cross section are given parameters.
Since the dark matter density is continuous for $r > r_{inner}$, and we assume that the dark matter density profile outisde the spike is an 
NFW profile with known parameters, the dark matter density profile inside the spike is determined by only one 
additional parameter.  We take this parameter to be the spike radius.  The dark matter density at $r_{sp}$ is then 
determined by matching to the NFW halo profile.  The density then increases with slope $\gamma_{sp}$ as $r$ decreases until 
$\rho (r_{core}) = m/\langle \sigma v \rangle t_{BH}$, which determines $r_{core}$.

One approach to determining the spike radius, often followed 
for the Milky Way, is to take $r_{sp}\approx 0.2 r_{h}$, where $r_h$ is the 
radius of gravitational influence of the black hole. 
This choice is motivated by results from numerical simulations of spike regeneration following a 
black hole merger~\cite{Merritt:2005yt}, and we will use this as a somewhat conservative benchmark.
Following~\cite{Shelton:2015aqa}, we assume that  the radius of gravitational influence 
satisfies the relation $r_h = G_N M_{BH} / \langle v^2 \rangle_{stellar}$, where
$\langle v^2 \rangle_{stellar}$ is the stellar velocity dispersion in the vicinity of the SMBH.  
Observations suggest $[\langle v^2 \rangle_{stellar}]^{1/2} \sim 420~\km / \s$, yielding 
$r_h \sim 150~\pc$ and $r_s \sim 30~\pc$~\cite{Murphy:2014eya}.

An alternative approach, used in~\cite{Gorchtein:2010xa,LBS:2015}, is to set an upper bound on the dark matter density in the spike 
by requiring that the dark matter contained within the radius of influence equals the uncertainty in the 
black hole mass.  This approach was applied to M87 in reference~\cite{LBS:2015},  which adopted the 
estimates $r_h = 10^5 r_{sch} \sim 60~\pc$, and $\Delta M_{BH} = 5 \times 10^8 M_\odot$~\cite{Gorchtein:2010xa}, 
yielding $r_{sp} = 220~\pc$.\footnote{We adopt $r_{sp} = 220~\pc$ for this approach, even 
though Ref.~\cite{LBS:2015} uses a smoothed version of our profile, since the resulting difference 
is small compared to the other uncertainties in this approach.}  Note that the gravitational radii of influence 
found by both approaches are roughly in agreement.  But the first approach leads to a spike radius which is a 
factor of $\sim 7$ smaller than the second.  The second approach, which may be thought of as an upper bound 
on the size of the dark matter spike, will thus result in a significantly larger gamma-ray flux than the 
first approach.

\subsection{$p$-wave annihilation within the spike}
\label{sec:p_wave}

For the case of $p$-wave annihilation, we assume that the dark matter annihilation 
cross section can be written as $\sigma v = (\sigma v)_0 (v/c)^2$, where $v$ is the 
relative velocity.  
This form of the annihilation cross section can arise in a variety of well-motivated 
theoretical models, including, for example, the annihilation of Majorana fermion dark matter 
to Standard Model (SM) fermion/anti-fermion pairs~\cite{Kumar:2013iva}.

The velocity-dependent form of the cross section will have two major 
effects on the spike $J$-factor, as compared to the $s$-wave annihilation case: it will 
change the photon flux produced by dark matter annihilation, and it will change the core 
radius by altering the conditions under which annihilation depletes the spike.

For the form of the density profile which we assume, the velocity dispersion was calculated 
in~\cite{Shelton:2015aqa} using the spherical Jeans equation, finding
\bea
\langle v^2 \rangle (r) &=& \frac{1}{c(r)} \frac{G_N M_{BH}}{r} ,
\eea
where $c(r)$ is a number which varies between 1 and $1+\gamma_{sp}$.  For simplicity, we will set 
$c(r)=1$.  

Using this expression, and averaging over the core~\cite{Boucher:2021mii} (assuming\footnote{Note, the approximation $r_{inner} \ll r_{core}$ will not be valid in the case where the 
core is very small, but this regime will not be relevant for our subsequent analysis.} $r_{inner} \ll r_{core}$), we find 
\bea
\overline{\langle \sigma v \rangle}_{n=2} &=&
\left(\frac{4\pi}{3} r_{core}^3 \right)^{-1}
\left[(2)4\pi \int_0^{r_{core}} dr~r^2~(\sigma v)_0 ~\langle v^2 \rangle(r) /c^2 \right] ,
\nonumber\\
&=& \frac{6 G_N M_{BH}}{c^2 r_{core}} (\sigma v)_0 \int_0^1 dx~ x
= \frac{3 G_N M_{BH}}{c^2 r_{core}} (\sigma v)_0 .
\eea
$\rho_{core}$ and $r_{core}$ are then determined by the relation
$(\rho (r_{core})/m) \overline{\langle \sigma v \rangle}_{n=2}  t_{BH} =1$.
Given choices for $r_{sp}$, $\gamma_{sp}$, $m$ and $(\sigma v)_0$, the density profile is 
now determined.  See Fig.~\ref{fig:profiles} for examples.

Assuming that the dark matter particle is its own antiparticle, 
the photon flux due to dark matter annihilation in the spike can be written as 
\bea
\frac{d\Phi_\gamma}{dE_\gamma} &=& \frac{d \Phi_{PP}}{dE_\gamma} \times J_{p}^{spike} ,
\eea
where $J_p^{spike}$ is the total $p$-wave $J$-factor for the spike, and 
\bea
\frac{d \Phi_{PP}}{dE_\gamma} &=& \frac{(\sigma v)_0}{8\pi m^2} \frac{dN_\gamma}{dE_\gamma} .
\eea
Here $m$ is the dark matter mass and $dN_\gamma / dE_\gamma$ is the photon spectrum per 
annihilation.  $d \Phi_{PP} / dE_\gamma$ is dependent only on particle physics properties, 
and is independent of the astrophysics of the spike.

We can express the total $J$-factor for $p$-wave annihilation within the dark matter spike as 
\bea
J_{p}^{spike} &=& \frac{1}{D^2} \int_{r_{inner}}^{r_{sp}} d^3 r \int d^3 v_1 \int d^3 v_2~ 
f(\vec{r}, \vec{v}_1)~f(\vec{r}, \vec{v}_2)~ (|\vec{v}_1 - \vec{v}_2|/c)^2 ,
\nonumber\\
\eea
where $D = 16~\mpc$ is the distance to M87~\cite{{Bird:2010rd}}, $f(\vec{r}, \vec{v})$ is the dark matter velocity
distribution within the spike, and we have used the fact that $r_s \ll D$. 
Assuming spherical symmetry and isotropy, we can express this as integral as~\cite{Boucher:2021mii}
\bea
J_{p}^{spike} &=& \frac{8\pi}{D^2} \int_{r_{inner}}^{r_{sp}} dr~ r^2 \rho^2(r) 
\frac{\langle v^2 \rangle(r)}{c^2} ,
\nonumber\\
&=& \frac{8\pi G_N M_{BH}}{c^2 D^2} \int_{r_{inner}}^{r_{sp}} dr~ r \rho^2(r) .
\eea

\begin{figure}[h]
\centering
\includegraphics[width=\textwidth]{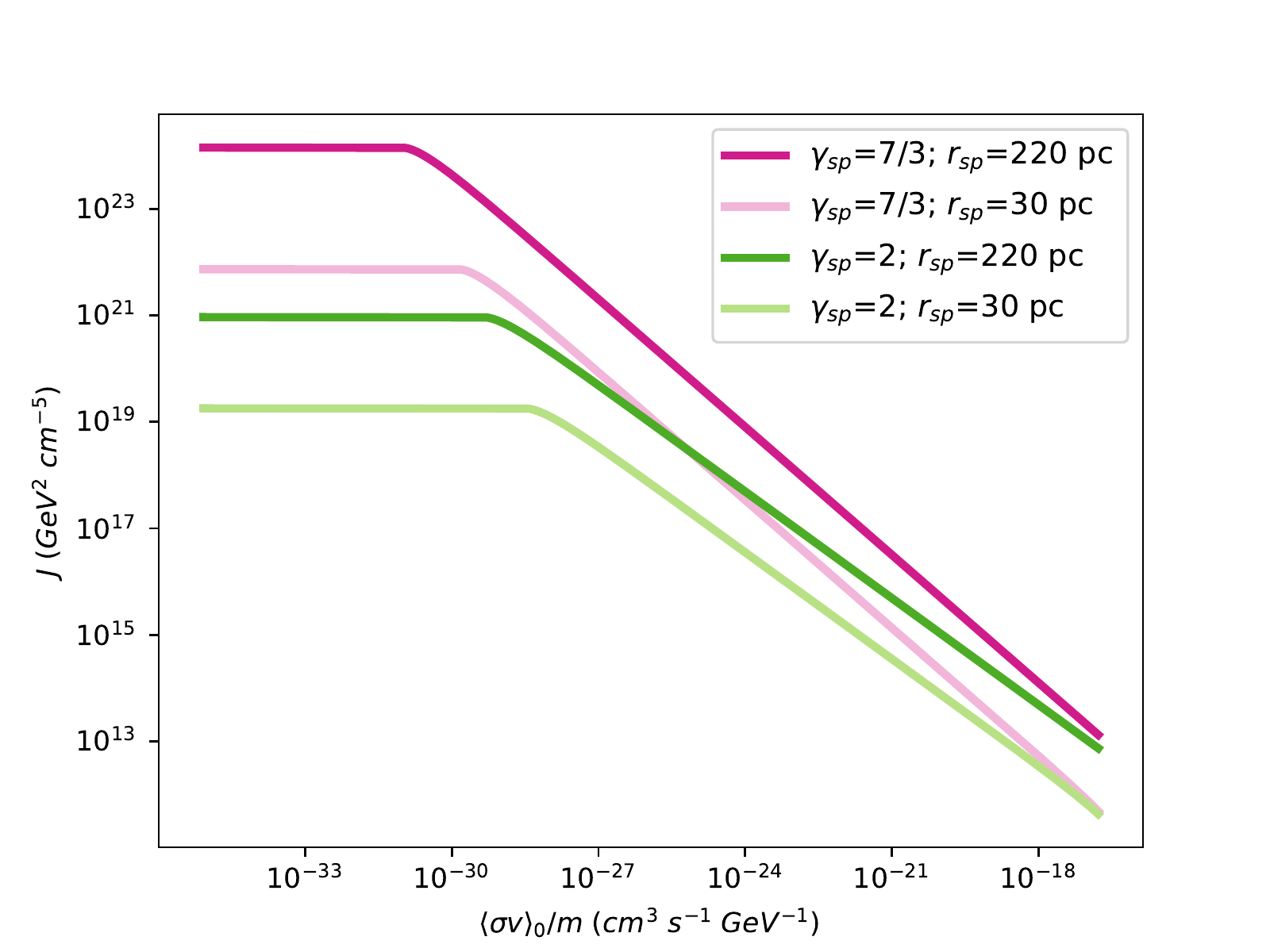}
\caption{The total $J$-factor for $p$-wave annihilation within the spike ($J_p^{spike}$) as a function of 
$(\sigma v)_0/m$, for $r_{sp} = 30~\pc,~ 220~\pc,$ and 
$\gamma_{sp} = 7/3$, $2$, as labeled.}
\label{fig:JvsigDM}
\end{figure}

We now have an expression for the total $p$-wave $J$-factor of the spike which requires only the density 
profile.  Using the general form of the profile which we have adopted, the spike $J$-factor is then 
entirely determined by $(\sigma v)_0 / m$ (the combination of dark matter particle physics parameters which 
determines the core radius), as well  
$r_{sp}$ and $\gamma_{sp}$ (note that we fix $r_{inner} = 4r_{sch}$).  
We consider the two motivated choices of $r_{sp}$ ($30~\pc$ and $220~\pc$) as discussed above.  We  also consider two 
choices for $\gamma_{sp}$: $\gamma_{sp} =7/3$, as would be expected from eq.~\ref{eq:gamma_sp} for an 
undepleted spike with $\gamma_c =1$, and a shallower choice, $\gamma_{sp} =2$.  
Note that for the shallower choice of $\gamma_{sp}$, the two approaches to fixing  $r_{sp}$ discussed in Sec.~\ref{sec:spikeprofile} yield 
slightly different results.  As this will not affect the $J$-factor significantly, we ignore this 
effect for simplicity.  For these choices of $r_{sp}$ and $\gamma_{sp}$, we plot the total $p$-wave  
$J$-factor for the M87 spike as a function of $( \sigma v)_0 / m$ in Fig.~\ref{fig:JvsigDM}.  The 
plateaulike features in the $J$-factor occur at the value of $(\sigma v)_0 / m$ at which $r_{core} = r_{inner}$.
For smaller values of $(\sigma v)_0 / m$, the dark matter density is not depleted appreciably by annihilation, 
and the $J$-factor is independent of $(\sigma v)_0 / m$.

For almost 
the entire range of parameters, including those of most interest, we find that the $p$-wave $J$-factor of the 
spike by far exceeds that of the rest of the halo.  Essentially, one can ignore the rest of the halo, 
and focus only on the dark matter spike (for the case of $s$-wave annihilation, this was already found 
in~\cite{LBS:2015}).  This stands in contrast to the more commonly-studied case of 
$s$-wave annihilation near Sgr A*, which is expected to produce a luminosity which is only a fraction of that 
of the entire Milky Way halo.  The difference is that the SMBH at the center of M87 is much larger than 
Sgr A*, leading to a much larger spike radius.  This effect is even more significant for the case of 
$p$-wave annihilation, because the velocity dispersion within the spike will be much larger.

The dependence of the gamma-ray flux on $(\sigma v)_0$ and $m$ is more complicated than 
one would expect in spikeless halos because of the depletion of dark matter within 
the core due to annihilation.  Outside the core (but within the spike), the rate of $p$-wave 
dark matter annihilation per radial shell is
$d\Gamma / dr \propto \left((\sigma v)_0 /m^2\right) r^{1-2\gamma_{sp}}$.  For the relatively large values 
of $\gamma_{sp}$ which we consider, the gamma-ray flux is primarily generated close to the core radius, 
yielding a total annihilation rate~\cite{Shelton:2015aqa}
\bea
\label{eq:rate}
\Gamma &\propto& \left( \frac{(\sigma v)_0}{m^2} \right) r_{core}^{2-2\gamma_{sp}} 
\propto \frac{(\sigma v)_0^{(3-\gamma_{sp})/(\gamma_{sp}+1)}}{(m^2)^{2/(\gamma_{sp}+1)}}.
\eea
To obtain Eq.~(\ref{eq:rate}), we have used the fact that, for fixed $r_{sp}$, $r_{core} \propto \left( (\sigma v)_0/m^{ }\right)^{1/(\gamma_{sp}+1)}$, 
which implies that the $J$-factor scales as 
\bea
J \propto \left(\frac{(\sigma v)_0}{m}\right)^{(2-2\gamma_{sp})/(\gamma_{sp}+1)}, 
\eea
which is approximately the behavior seen in Fig.~\ref{fig:JvsigDM}.
As expected, the annihilation rate [Eq.~(\ref{eq:rate})] increases with increasing $(\sigma v)_0$ 
(the normalization of the annihilation cross section) and with decreasing $m$ (which increases 
the number density).  Neither of these effects is as pronounced in a depleted halo as it would be in a halo that is undepleted by 
dark matter annihilation, where one expects $\Gamma\propto (\sigma v)_0/m^2$.  Instead, increasing $(\sigma v)_0$ or decreasing $m$ also increases 
the core radius, inside which the density has saturated.  Note, though, that this simple scaling 
relation will not entirely determine the shape of the exclusion contour derived from gamma-ray data, 
since the gamma-ray spectrum per annihilation also depends on $m$.

Assuming $\gamma_c =1$, then in the case of $\gamma_{sp} = 7/3~(2)$, changing the spike radius 
from $220~\pc$ to $30~\pc$ rescales the density in the region $r < 30~\pc$ by the factor 
$0.7~(0.14)$.  This in turn leads to a rescaling of the core radius by a factor of $\sim 0.45~(0.51)$, 
yielding a rescaling of the annihilation rate in the spike  (outside the core) of roughly 
$0.04~(0.07)$.  We thus expect that the most optimistic choice of spike radius would lead to a gamma-ray 
flux enhanced by roughly a factor $15 -25$ over a choice motivated by simulations.

\section{Results}
\label{sec:results}

Here we use Fermi-LAT~\cite{Fermi-LAT:2009ihh} Pass 8 and MAGIC~\cite{Magic_the_Gathering:2020} data 
in the $1~\gev-10~\tev$ range, as reported in~\cite{Magic_the_Gathering:2020}, to constrain dark matter annihilation in a dark matter spike in M87.
Of course, dark matter annihilation in a galactic 
environment can produce gamma rays outside this energy range, as well as x-rays due, for example, to 
synchrotron radiation from charged annihilation products.  Thus, many other datasets can be 
used to constrain dark matter annihilation in M87.  That said, we find that Fermi-LAT and MAGIC data alone 
can provide interesting constraints.  We leave the application of this formalism to other 
datasets for future work.

We take the conservative perspective that all gamma-rays observed from M87  
are due to dark matter annihilation.  A model is excluded if it would yield an expected 
number of photons in any energy bin which exceeds that observed in Fermi-LAT or MAGIC data by more than 
$1\sigma$.

\begin{figure}[h]
\centering
\begin{subfigure}[b]{0.49\textwidth}
\centering
\includegraphics[width=\textwidth]{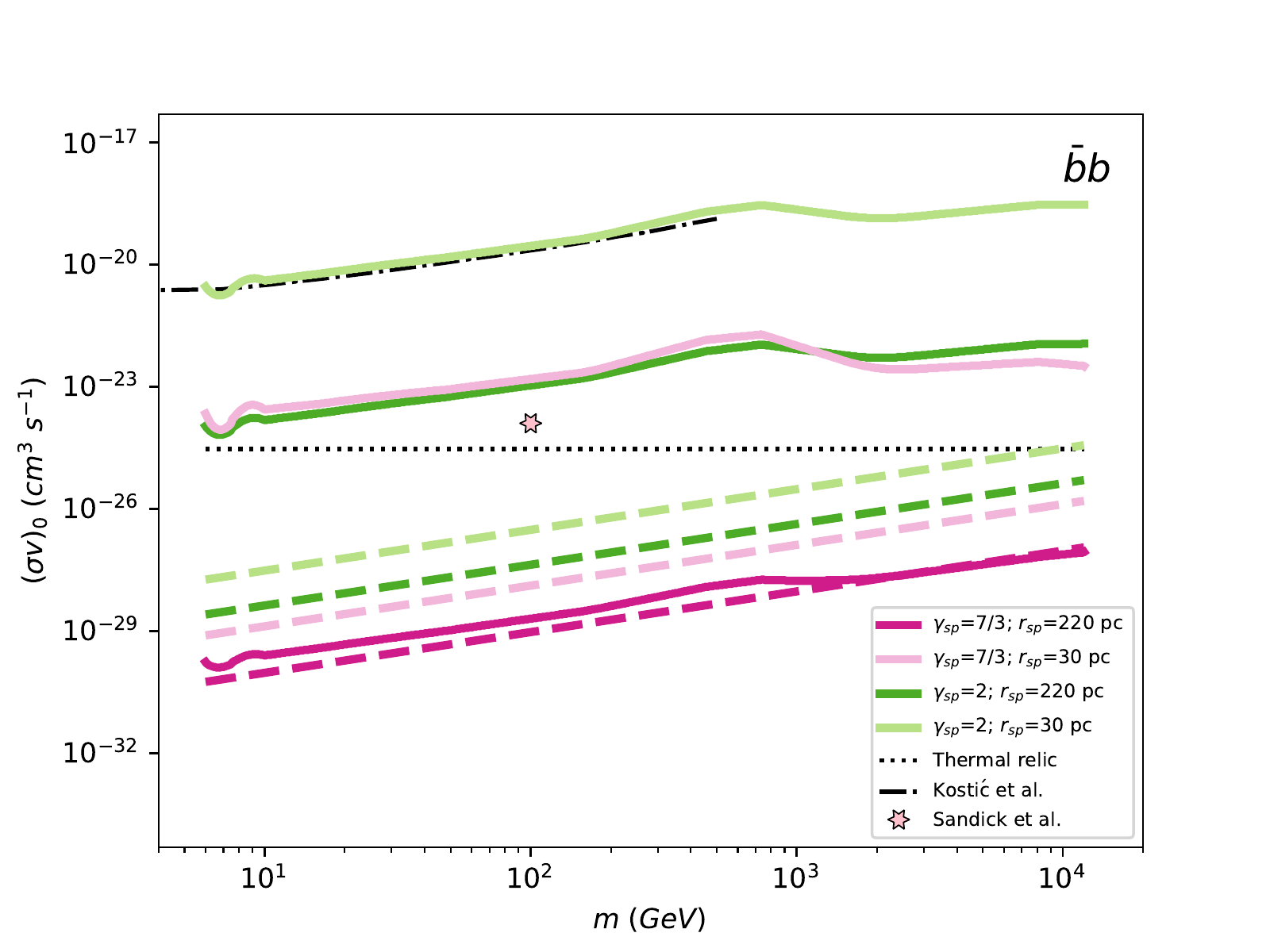}
\end{subfigure}
\begin{subfigure}[b]{0.49\textwidth}
\centering
\includegraphics[width=\textwidth]{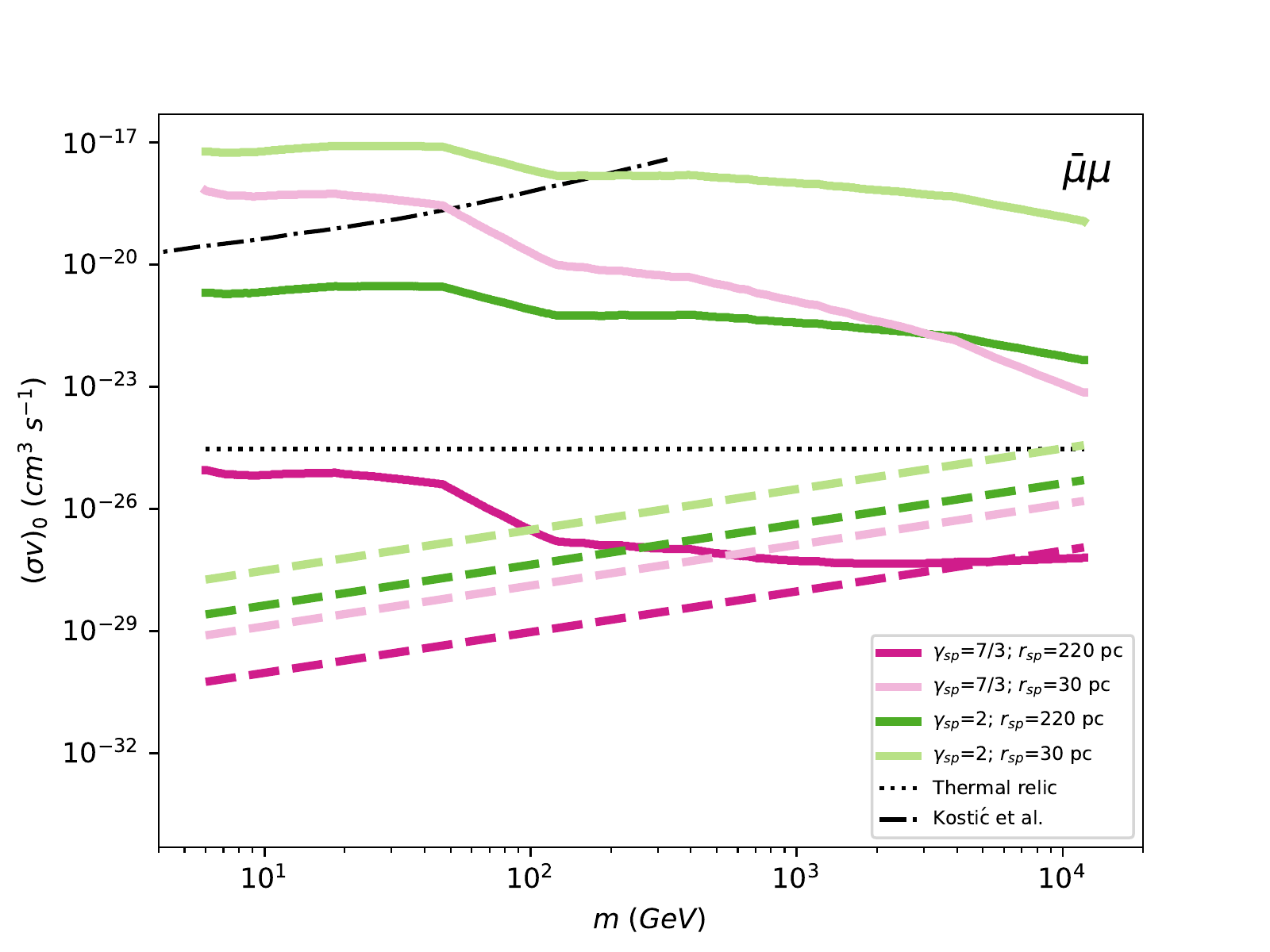}
\end{subfigure}
\caption{Exclusion contours (solid lines) in the $(m,(\sigma v)_0)$-plane, assuming dark matter annihilates entirely to 
$\bar b b$ (left panel) or $\bar \mu \mu$ (right panel).  We take $\gamma_{sp} = 7/3,~2,$ and $r_{sp} = 220~\pc,~30~\pc$, 
as indicated.  Dashed lines are contours of constant $(\sigma v)_0/m$ for which $r_{core} = r_{inner}$.  The dotted gray 
line indicates the value of $(\sigma v)_0$ for a $p$-wave thermal relic.  The dot-dashed black lines are constraints from 
Kostic et al.~\cite{Kostic:2023arx}, as discussed in the text.
}
\label{fig:cslimits}
\end{figure}

For simplicity, we consider two annihilation channels: $\bar b b$, which tends to yield 
a relatively large number of high-energy photons per annihilation, and $\bar \mu \mu$, which tends to 
yield a small number of high-energy photons.  For both channels, the photon spectrum per annihilation was 
obtained from~\cite{PPPC4DMID:2011}.
As discussed in Sec.~\ref{sec:formalism}, we consider 
the cases $\gamma_{sp} = 7/3, 2$ and the cases $r_{sp} = 30~\pc, 220~\pc$.  
In Fig.~\ref{fig:cslimits}, we present exclusion contours in the 
$\left(m, (\sigma v)_0\right)$-plane for annihilation to $\bar b b$ (left panel) and $\bar \mu \mu$ (right panel).  
In both panels, the gray dotted line indicates the value of $(\sigma v)_0$ for which 
the relic density can be explained by thermal freeze-out through $p$-wave annihilation, i.e.~$(\sigma v)_0 \sim 
3 \times 10^{-25}~\cm^3/\s$, with $\langle v^2/c^2 \rangle \sim 0.1$.   
As an initial matter, we note that in all cases we have considered, the 
constraints obtained from M87 are stronger than those obtained by observations of dSphs~\cite{Boddy:2019qak,Boddy:2019kuw}, 
which do not appear on the scale plotted.  Moreover, for much of the parameter space, the constraints obtained from M87 
surpass those obtained from a recent 
search for $p$-wave annihilation in local large scale structure~\cite{Kostic:2023arx}, denoted as dot-dashed black lines in Fig.~\ref{fig:cslimits}.

\begin{figure}[h]
\centering
\begin{subfigure}[b]{0.49\textwidth}
\centering
\includegraphics[width=\textwidth]{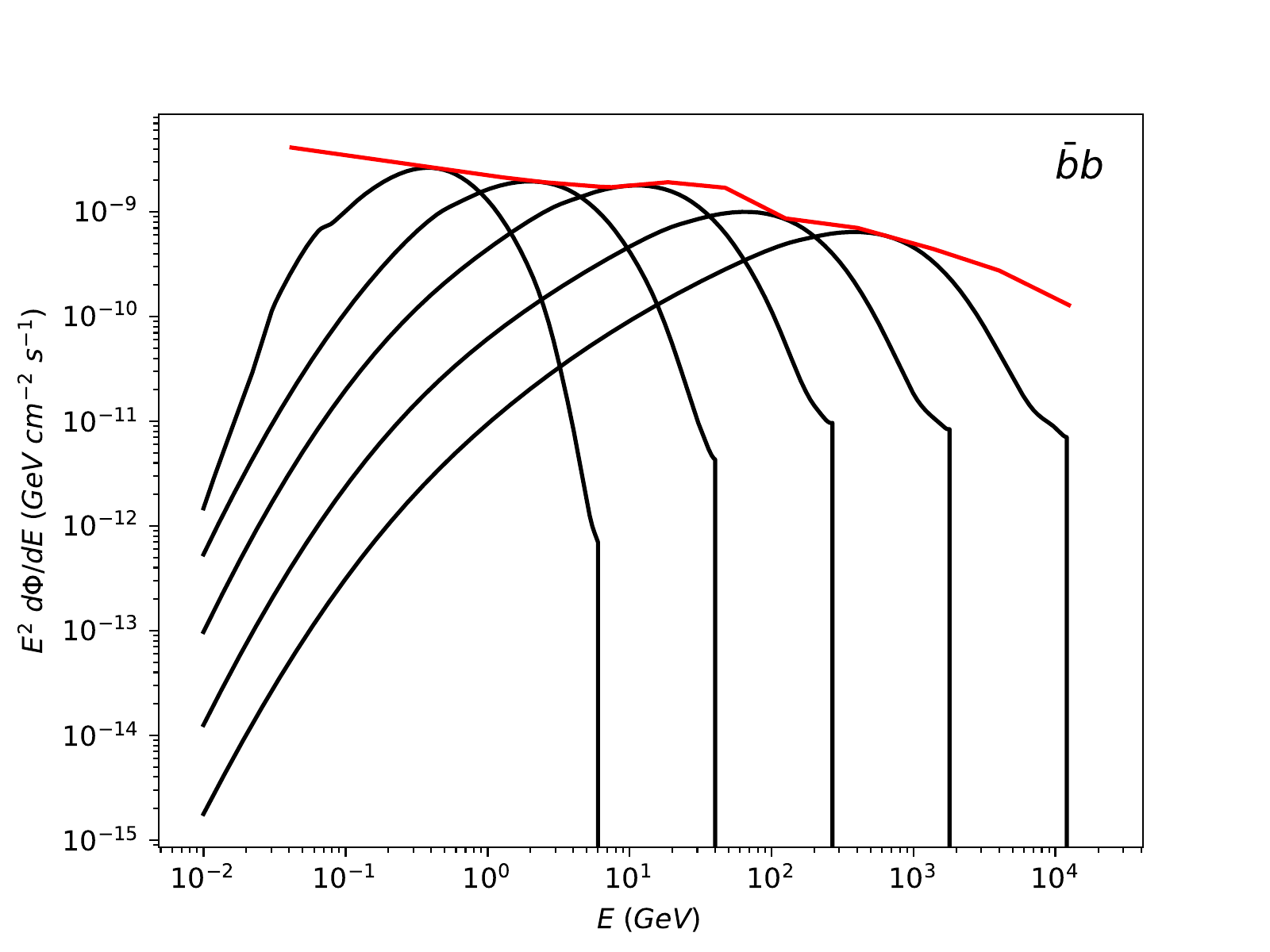}
\end{subfigure}
\begin{subfigure}[b]{0.49\textwidth}
\centering
\includegraphics[width=\textwidth]{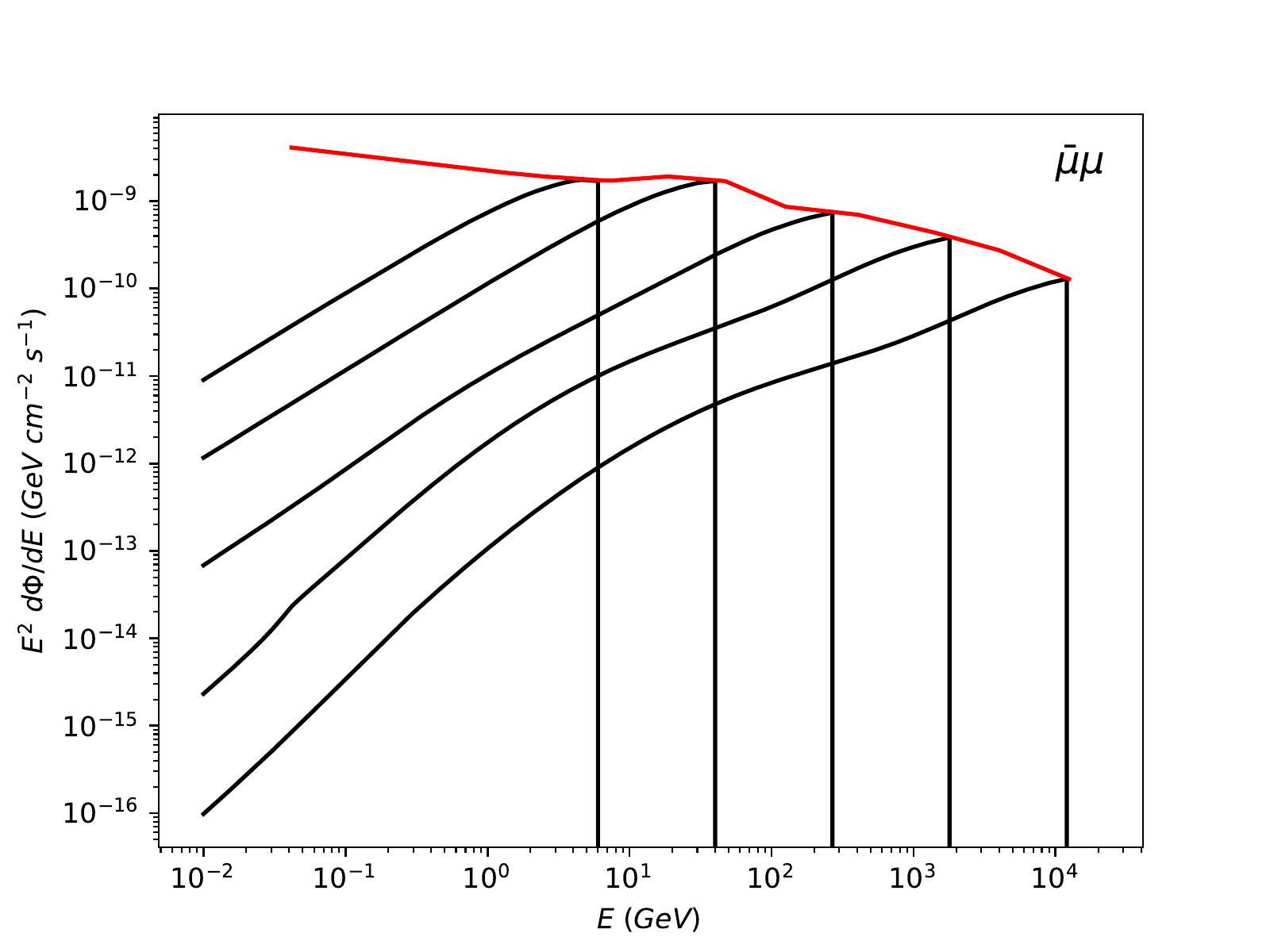}
\end{subfigure}
\caption{The differential photon flux $E^2 d\Phi / dE$ (black lines) produced by $p$-wave annihilation to 
$\bar b b$ (left panel) and $\bar \mu \mu$ (right panel) for m = ($6~\gev$, $40~\gev$, $268~\gev$, $1.8~\tev$, $12~\tev$) from left to right, assuming that 
$(\sigma v)_0$ is chosen to lie on the appropriate exclusion contour.  The red line in each panel is the differential 
photon flux observed by Fermi-LAT~\cite{Fermi-LAT:2009ihh} and MAGIC~\cite{Magic_the_Gathering:2020}, as reported 
in~\cite{Magic_the_Gathering:2020} (error bars are suppressed).}
\label{fig:fluxlimits}
\end{figure}

For the most optimistic case of $\gamma_{sp} = 7/3$, $r_{sp} = 220~\pc$, scenarios of thermal dark matter 
with $p$-wave annihilation are ruled out even for masses as large as $10~\tev$, regardless of the annihilation 
channel.  But for a smaller spike radius favored by simulations, or for a shallower spike slope, exclusion 
bounds are weakened dramatically. This is because a reduction in the spike slope or radius will 
reduce the gamma-ray flux dramatically (see discussion at the end of Sec.~\ref{sec:p_wave}), 
whereas in order to compensate for such a decrease in the flux, a much larger cross section would be required.
Dark matter annihilation is maximized near the core radius, where the dark matter density saturates.  The effect of 
increasing the annihilation cross section is not to increase the annihilation rate within the core, but 
rather to increase the size of the core.  As such, a large increase in the annihilation cross section is 
needed to achieve even a modest increase in the total annihilation rate.

It is difficult to make a direct comparison to the study in~\cite{Shelton:2015aqa} 
of $p$-wave annihilation in a dark matter spike around 
Sgr A*, since that work does not consider these annihilation channels, and presents limits for a continuum channel only 
at $m = 110~\gev$.  Nevertheless, \cite{Shelton:2015aqa} found some models of thermal $p$-wave dark matter could be ruled out, for 
$\gamma_{sp} = 7/3$, assuming a photon spectrum with no sharp features. 
We can compare our results more directly to the study in~\cite{Sandick:2016zeg}, which considered 
$p$-wave annihilation to the $\bar b b$ final state for $m = 100~\gev$, assuming  $\gamma_{sp}^{(Sgr A*)} = 7/3$ and 
$r_{sp}^{(Sgr A*)} = 0.4~\pc$ (using the relation $r_{sp} = 0.2 r_h$).  The bound found in~\cite{Sandick:2016zeg} is indicated with a star in 
Fig.~\ref{fig:cslimits}.  We see that constraints from M87 are much stronger, for $\gamma_{sp} = 7/3$  and an optimistic choice of the spike radius.  But the 
constraints from M87 for a spike radius of $30~\pc$ are somewhat weaker than from Sgr A* for the case of 
an undepleted spike.  But, as noted in~\cite{Sandick:2016zeg}, it is also quite possible that the DM spike 
around Sgr A* has been depleted by interactions with stars.  The timescale for the heating of DM via scattering 
off stars in 
Sgr A* is ${\cal O}(10^9)~\yr$~\cite{Bertone:2005hw}, about a factor of 10 shorter than the 
estimated age of Sgr A*.  Based on these 
estimates, depletion of the spike around Sgr A* could reduce the spike radius size by a factor of 
$\sim 4$~\cite{Sandick:2016zeg}.  The resulting photon flux, for a fixed  annihilation cross section, would be 
$1-2$ orders of magnitude smaller.  But since DM spike around Sgr A* is estimated to have a core~\cite{Sandick:2016zeg}, 
and increasing the annihilation cross section will increase the core size, the weakening of  bounds on the 
annihilation cross section due to depletion of the Sgr A* spike will actually be much more severe.
On the other hand, the timescale for heating of M87 is estimated to 
be ${\cal O}(10^{14})~\yr$~\cite{LBS:2015}, which is orders of magnitude larger than the age of the Universe, 
and implies that a similar depletion of the spike in M87 is much less likely.
One may conclude that, in comparing the SMBH 
at the center of M87 to that at the center 
of the Milky Way, as a target for dark matter searches, the preference for target is dominated by astrophysical 
uncertainties.
M87 is likely to be a better target if the spike radius is a large as observations allow, 
and/or if, as one might expect, its spike is undepleted by scattering with stars, while the spike around 
Sgr A* is depleted.

Also plotted in Fig.~\ref{fig:cslimits} are dashed lines of constant 
$(\sigma v)_0/m$ at which the core disappears, for the various choices of $r_{sp}$ and 
$\gamma_{sp}$.  Below these lines, dark matter annihilation does not saturate.  
We see that the exclusion contours all lie almost entirely in the 
region for which there is a core within which dark matter annihilation has saturated.

It is interesting to compare the behavior of the exclusion contours for the $\bar b b$ 
and $\bar \mu \mu$ channels.  As expected, the exclusion contours for the $\bar b b$ channel 
lie at smaller cross sections than those for the $\bar \mu \mu$ channel.  But, perhaps unexpectedly, 
we find for the $\bar \mu \mu$ channel that exclusion contours strengthen as the dark matter 
mass increases.  This is opposite to the behavior of the $\bar b b$ channel, and opposite to 
the usual expectation from searches of dark matter annihilation in halos.

In halos in which the dark matter density is not significantly depleted by annihilation, 
the total annihilation rate increases with cross section and decreases with mass as 
$(\sigma v)_0/m^2$.  In a spike in which depletion effects are significant, we have seen 
that this dependence is weakened.  But there is an additional dependence of the photon 
spectrum on the dark matter mass.  Because we consider a conservative analysis, a model is 
considered excluded if there is any energy bin in which the model predicts a flux which exceeds 
observation (within uncertainties).  Because the observed gamma-ray flux from M87 decreases 
with energy roughly as $d\Phi/dE \propto E^{-2.24}$~\cite{Magic_the_Gathering:2020,Fermi:2009}, 
the allowed flux due to dark matter annihilation decreases 
rapidly with increasing dark matter mass.  As a result, the exclusion contours for the 
$\bar b b$ channel weaken only slightly with increasing dark matter mass, while the contours 
for the $\bar \mu \mu$ channel strengthen.  

To illustrate this point, we plot in Fig.~\ref{fig:fluxlimits} the differential photon flux produced 
for the $\bar b b$-channel (left panel) and the $\bar \mu \mu$-channel (right flux) for 
$m = 6~\gev, 40~\gev, 268~\gev, 1.8~\tev$, and  $12~\tev$, assuming that $(\sigma v)_0$ is chosen to lie on 
the appropriate exclusion contour.  The red curve in both panels is the differential flux observed by 
Fermi-LAT and MAGIC.

\section{Conclusion}
\label{sec:conclusion}

We have considered the prospects for constraining scenarios of $p$-wave dark matter annihilation 
with observations of M87.  The SMBH at the core of M87 is an interesting target because, though very 
far away, it is about 1000 times more massive than the SMBH at the center of our own galaxy.  As a 
result, it may be surrounded by a very dense spike of very fast-moving dark matter particles, for 
which $p$-wave annihilation is enhanced.

We have considered a conservative data analysis, in which we assume that all gamma-rays arriving from 
M87 are due to dark matter annihilation in the spike.  
If the spike forms by adiabatic contraction and is not disrupted by galactic dynamics, then the bounds 
obtained from Fermi-LAT and MAGIC data are stronger than current bounds on $p$-wave annihilation in 
dwarf spheroidal galaxies.
We find that if the dark matter spike is particularly 
fortuitous (that is, with a very steep slope and a size as large as is allowed by observation), 
then current observations can rule out a thermal $p$-wave dark matter candidate by several orders of magnitude, even 
for a final state such as $\bar \mu \mu$ which produces relatively few photons per annihilation.  On the 
other hand, if the spike is smaller 
or less steep, then scenarios 
of thermal $p$-wave dark matter are essentially unconstrained.  The reason is that, in order to obtain a large
signal, the dark matter annihilation cross section must usually be large enough that dark matter is depleted 
in the innermost regions, forming a core.  However, if this signal is not large enough to completely explain the 
observed flux, then it is very difficult to increase the flux by increasing the cross section, since this would 
further deplete the core.

One can see that exclusion limits on dark matter matter models depend very strongly on how large a gamma-ray 
flux can be accommodated by the data.  We have used a very conservative analysis, in which no attempt is made 
to model astrophysical backgrounds, and all photons from M87 are assumed to arise from dark matter annihilation.
In other words, the gamma-ray flux which can be attributed to dark matter annihilation is as large as possible.  But there 
is expected to be a large flux of gamma-rays arising from astrophysical processes, such as jets produced in the 
vicinity of the SMBH.  If these backgrounds can be modelled, then the flux potentially attributable to dark matter 
annihilation would be reduced.  This would strengthen bounds on dark matter annihilation substantially, for the 
reasons described above.

In this work, we have only considered the gamma rays produced promptly by dark matter annihilation, constrained 
by data from Fermi-LAT and MAGIC.  But processes such as synchrotron radiation can copiously produce lower energy 
photons, which can potentially be constrained even more tightly by other datasets~\cite{LBS:2015}.  We have 
not considered these constraints because they depend on a variety of additional systematic uncertainties, such 
as the magnetic fields near the center of M87.  But a more detailed study of these approaches is warranted.

For the case of $p$-wave annihilation, if the dark matter spike forms by adiabatic contraction and is undepleted, then the luminosity of the spike may easily dominate that of the rest of the halo.  
Of course, it is quite possible that the spike is depleted by galactic dynamics.  In any case, 
this indicates the level of systematic uncertainty in the dark matter annihilation signal.  Essentially, 
even complete knowledge of the M87 dark matter halo profile outside the spike tells us virtually nothing about the 
luminosity due to dark matter annihilation, unless we also have knowledge of the spike parameters.  This 
is largely a function of the size of the SMBH in M87.  Indeed, M87 hosts an AGN, and thus is expected to 
provide a large background of gamma-rays sourced by astrophysical processes in the vicinity of the SMBH, 
in addition to any potential gamma-rays from the dark matter spike.  Recent work in the literature considers the 
possibility of searching for dark matter annihilation in extragalactic halos (see, 
for example,~\cite{Baxter:2022dpn,Kostic:2023arx}), 
for which correlated 
astrophysical backgrounds such as AGNs, are a major difficulty which must be addressed.  Dark matter annihilation 
in the spike can also provide a signal which cannot be simply correlated to the halo parameters.

Finally, we note that, although we have focused on the case of $p$-wave annihilation, many similar considerations 
will hold for the case of $d$-wave annihilation, in which the annihilation cross section scales as 
$\propto (v/c)^4$.

\centering{\bf ACKNOWLEDGEMENTS}  

We are grateful to Celine Boehm and Jonelle Walsh for useful discussions.  JK is supported in part by 
DOE grant DE-SC0010504.  PS is supported in part by NSF grant PHY-2014075.  JK is grateful to the 
University of Utah, where part of this work was done, for its hospitality.

\end{document}